# Near-Field UHF RFID Transponder with a Screen-Printed Graphene Antenna

Kaarle Jaakkola, Henrik Sandberg, Markku Lahti and Vladimir Ermolov

*Abstract*—As a method of producing RFID tags, printed graphene provides a low-cost and eco-friendly alternative to the etching of aluminum or copper. The high resistivity of graphene, however, sets a challenge for the antenna design. In practice, it has led to using very large antennas in the UHF RFID far field tags demonstrated before. Using inductive near field as the coupling method between the reader and the tag is an alternative to the radiating far field also at UHF. The read range of such a near field tag is very short, but, on the other hand, the tag is extremely simple and small.

In this paper, near field UHF RFID transponders with screen-printed graphene antennas are presented and the effect of the dimensions of the tag and the attachment method of the microchip studied. The attachment of the microchip is an important step of the fabrication process of a tag that has its impact on the final cost of a tag. Of the tags demonstrated, even the smallest one with the outer dimensions of 21 mm * 18 mm and the chip attached with isotropic conductive adhesive (ICA) was readable from a distance of 10 mm with an RF power marginal of 19 dB, which demonstrates that an operational and small graphene-based UHF RFID tag can be fabricated with low-cost industrial processes.

*Index Terms*— antenna, graphene, inductive near field, ink, integration, isotropic conductive adhesive (ICA), near field UHF RFID, printing, RFID

## I. INTRODUCTION

RADIO-FREQUENCY identification (RFID) is based on exploiting electromagnetic fields to transfer data wirelessly, for the purposes of automatically identifying and tracking tags attached to objects. The technology finds applications in many areas: supply chain management, inventory tracking, contactless payment etc. [1] [2] [3]. The elements of an RFID systems are transponders and a reader, which exchange information wirelessly. A passive RFID transponder consists of a microchip and an antenna. Ultra-high frequency (UHF, 860-960 MHz) RFID transponders with printed graphene antennas and read distances comparable with commercial RFID transponders have been demonstrated recently [4] [5]. Printed graphene-based antennas have several advantages such as low cost, chemical stability, mechanical flexibility, resistance against fatigue and eco-friendliness. For comparison, etching of metal is an environmentally unsafe process and the cost of silver ink strongly depends on the price level of bulk silver metal. The price of silver is very stable and is not expected to decrease in the foreseen future. The cost of graphene ink raw material is very low and the manufacturing process is simple and scalable. The cost can be estimated based on the complexity and scalability of the process and is expected to be much lower than that of commercially available silver inks.

UHF RFID transponders operate typically in the far field mode with electromagnetic waves propagating between reader and transponder antennas. This operation mode is optimal for applications for which long read distances (up to 10 meters) are required. However, there are applications in which RFID transponders do not need a long read range, for example tracking of small objects such as gadgets, pharmaceutical packages, bottles, cartridges or battery packs. For such applications, small size of a transponder is expected and the read range of under 0.5 meters can be accepted. Near field low frequency (LF, 125-134 KHz) and high frequency (HF, 13.56 MHz) RFID systems based on inductive coupling between reader and transponder antennas are commonly used in such applications today.

It has been demonstrated that the near field coupling can be utilised in the UHF frequency range as well [6]. In this case, magnetic (inductive) coupling or electric (capacitive) coupling can be employed between the antennas of a reader and a transponder. The utilization of near field UHF transponders allows the usage of a single UHF infrastructure for all needed RFID applications instead of the combination of UHF, LF and HF RFID systems. It saves the investments in tracking infrastructures considerably.

In this work, we present operational near field UHF RFID transponders with printed graphene antennas. The transponder is based on inductive coupling between the reader and the tag antennas. Because of the stronger inductive coupling at higher frequencies, magnetic UHF transponder has a single-loop antenna, which is much simpler and cheaper compared to LF/HF multi-turn antenna coils with a bridge between the first and the last turn. For mass-produced transponders in disposable applications, direct printing of carbon-based conducting materials is a sustainable and low-cost alternative, and the graphene-based printing ink offers a robust system with a reasonable conductivity.

Manuscript submitted Oct 20, 2018; revised Jan 10, 2019; accepted Feb 21, 2019. This work has been part of the Graphene Flagship project, which has received funding from the European Union's Horizon 2020 research and innovation programme under grant agreement No 785219. The partial funding from VTT Technical Research Centre of Finland Ltd is also gratefully acknowledged.

The authors are with the VTT Technical Research Centre of Finland Ltd., FI-02044 Espoo, Finland (e-mail: firstname.surname@vtt.fi).

## II. Design

In practice, the inductive near field transponders are readable from very close distance also with far field antennas, which is due to the magnetic near field component of any practical antenna, but in order to achieve an ideal read range, special near field antennas should be used [6]. Even though referred commonly as "antennas", the near field antennas are not, in fact, antennas, but inductors. Instead of radiating, their purpose is to form a high-frequency magnetic field that is, at least close to the antenna, as uniform as possible. In practice, both the reader and transponder inductors at UHF frequencies are single-turn planar coils and therefore referred to as "coils" in the following. Electromagnetically, the reader coil and the transponder coil form a transformer with the power coupling efficiency $H$. If we assume that the impedance of the reader coil is conjugately matched to the RF output of the reader and that the transponder antenna is reactively matched to a chip, the coupling efficiency can be calculated [7] [8]:

$$H = \frac{4k_c^2 Q_r Q_t \frac{R_{IC}}{R_t}}{\left[2\left(1+\frac{R_{IC}}{R_t}\right)+k_c^2 Q_r Q_t\right]^2}, \quad (1)$$

where $k_c$ is the coupling factor between the coils, $Q_r$ and $Q_t$ are the Q values of the reader and the transponder coils, respectively. $R_{IC}$ is the series resistance of the microchip and $R_t$ is the series resistance of the transponder coil. The coupling factor $k_c$ can be determined by the mutual inductance between the coils ($M$) and the self-inductances of the reader and the transponder coils $L_r$ and $L_t$:

$$k_c = \frac{M}{\sqrt{L_r L_t}}. \quad (2)$$

By simulations or measurements, $k_c$ can also be determined by using the inductance value of the primary coil in two cases: with the secondary coil shorted and the secondary coil left open. Using this method, $k_c$ becomes [7]:

$$k_c = \sqrt{1-\frac{L_s}{L_o}}, \quad (3)$$

where $L_s$ is the inductance of the primary coil with the secondary coil shorted and $L_o$ the inductance of the primary coil with the secondary coil left open.

As the transponder is a simple single-turn coil that is small compared to the wavelength, its Q value is equal to its inductive reactance divided by its series resistance:

$$Q_t = \frac{X_t}{R_t} = \frac{2\pi f L_t}{R_t}, \quad (4)$$

where $X_t$ is the inductive reactance of the coil, $R_t$ is the serial resistance of the coil and $f$ is frequency. Even though (1) defines the exact coupling efficiency, a more simple formula that is also a term of (1) can be used to define the figure of merit $U$ for an inductively coupled system [8] [9]:

$$U = k_c\sqrt{Q_r Q_t}, \quad (5)$$

Additionally to the coupling factor and the Q values, the third term affecting the coupling efficiency is the impedance match of both the reader coil and the transponder. The reader coil is designed to provide a good match to the 50-Ohm RF output of the reader. The coil on the transponder side is designed to match the reactance of the microchip, but due to its simple structure, the real parts of the impedances are typically not matched.

The reactive matching on the transponder side is done by adjusting the circumference of the loop until its reactance has the same but opposite value as that of the microchip. For near field transponders with an aluminum antenna, the real part of the impedance remains lower than that of the microchip, but in the case of materials with higher resistivity, such as graphene, the real part can be significantly higher. This means that the Q value of the transponder as a coil remains low, which reduces the coupling efficiency as defined by (1).

The transponder was designed with Ansys HFSS electromagnetic simulation tool. The simulation model shown in Fig. 1 comprises the transponder (gray, on top) and the reader coil. Impinj Matchbox™ was selected as the reader coil [10]. Matchbox contains a single-turn coil that is divided into four segments that are connected to each other with series capacitors. The capacitors shift the phase of the signal so that the structure, the circumference of which is already significant compared to the wavelength, produces a uniform magnetic field without radiating [11]. The simulation model shown in Fig. 1 contains two 50-Ohm lumped ports, of which port 1 is connected to the input of the transponder coil and port 2 is connected to the input of the reader coil. The power coupling $F$ between the reader coil and the RFID microchip can then be calculated using the simulated S parameters of the ports: $S_{11}$ and $S_{21}$:

$$F = \frac{|S_{21}|^2}{(1-|S_{11}|^2)}(1-|\Gamma|^2), \quad (6)$$

where $\Gamma$ is the reflection coefficient at the RF input of the transponder and is calculated from the input impedance of the transponder coil $Z_t$ ($= R_t + jX_t$) and the RF impedance of the microchip $Z_{IC}$ ($= R_{IC} + jX_{IC}$):

$$\Gamma = \frac{Z_t - Z_{IC}^*}{Z_t + Z_{IC}}, \quad (7)$$

In order to evaluate only the power coupling efficiency between the coils ($A$), omitting the effect of the impedance of the microchip, a simplified form of (6) can be used:

$$A = \frac{|S_{21}|^2}{(1-|S_{11}|^2)}. \quad (8)$$

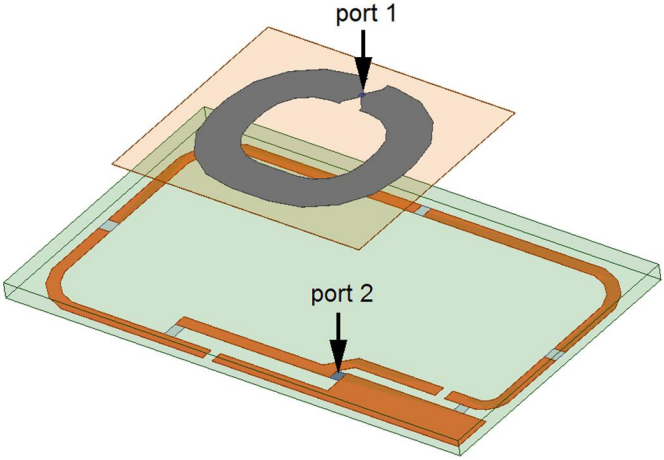

Fig. 1. Simulation model of the transponder and the reader coil.

The design guideline of the inductive near field transponder is based on the round shape that is optimal in terms of generating magnetic flux that protrudes out of the coil. As stated above, the circumference of the transponder coil is determined by the required reactance, which, in turn, is defined by the input impedance of the microchip. The input impedance of the selected microchip, Monza R6 by Impinj, is (13 - j126) Ohms at 867 MHz, making the ideal transponder input impedance (13 + j126) Ohms. The input impedance of Monza R6 includes the effect of the estimated stray capacitance due to adhesive bonding, as specified by Impinj [12]. The width of the conductor affects the resistance of the coil and gets in this case relatively large - several millimeters - due to the low conductivity of graphene. In order to find the right reactive tuning and to study the effect of the conductor width, prototypes with different dimensioning were fabricated. The effect of the square resistance was also studied. Figure 2 shows the critical dimensions of the transponder and the position of the microchip (IC). The reactance of the coil was adjusted to be close to 126 Ohms at 867 MHz by varying the dimension dx. The simulation results of the transponders with different parameter values are listed in Table 1; the input impedance of the transponder $Z_t$ calculated from $S_{11}$, the effect of the impedance mismatch as described by the term $1-|\Gamma|^2$ of (6) and the overall coupling factor $F$ as defined by (6) and (7).

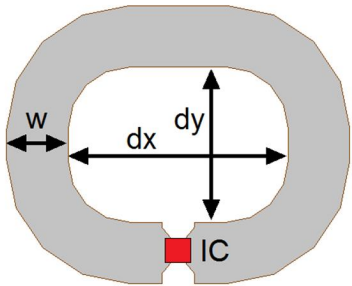

Fig. 2. Critical dimensions of the transponder prototypes.

Fig. 3 illustrates the simulated reflection coefficient $\Gamma$ as defined by (7) for the cases of Table I as a function of frequency. Fig. 3 shows that, even though the reactances are matched, due to the poor match between the real parts of the impedance, the reflection coefficient remains high and only the curve of the highest-Q case (5 Ohms 7 mm) has a mild form of resonance.

TABLE I
PARAMETER VALUES AND THE SIMULATION RESULTS OF THE TRANSPONDER COILS AT 867 MHZ.

| $R/\square$ ($\Omega$) | w (mm) | dx (mm) | dy (mm) | $Z_t$ ($\Omega$) | $1-|\Gamma|^2$ (dB) | F (dB) |
|---|---|---|---|---|---|---|
| 5 | 4 | 12.05 | 10.1 | 86 + j129 | -3.2 | -29.4 |
| 10 | 4 | 14.3 | 10.1 | 185+j129 | -5.9 | -34.4 |
| 5 | 7 | 11.8 | 10.1 | 70 + j129 | -2.6 | -25.7 |
| 10 | 7 | 12.8 | 10.1 | 136+j127 | -4.8 | -30.2 |

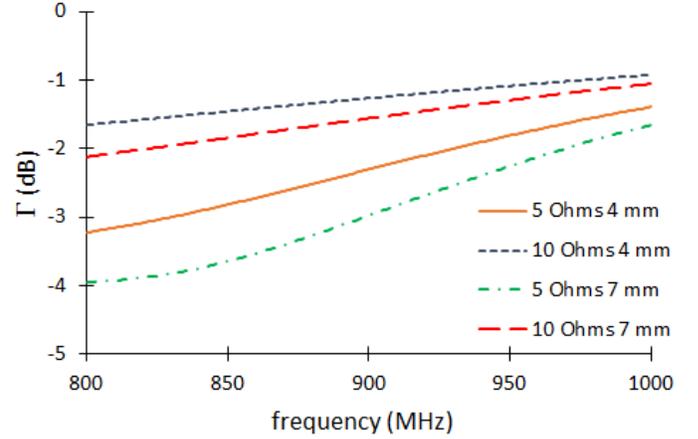

Fig. 3. Simulated reflection coefficient of the transponders of Table I as a function of frequency.

As the Matchbox reader antenna is, instead of a single inductor, a set of distributed inductors in series, unambiguous single inductance or respective Q value cannot be defined for it. However, the coupling factor $k_c$ between the coils can be determined using (3). In the simulations made to determine the values of $k_c$, the transponder coil is used as the measured primary coil and the input of Matchbox is shorted and left open, as explained above. Table II lists the values of $k_c$ obtained this way for the same cases as introduced in Table I. Additionally to $k_c$, the changing Q value of the transponder coil has its impact on the coupling efficiency. Therefore, the figure of merit as defined by (5) is studied here as well. As the Q value of Matchbox $Q_r$ is undefined but constant, Table II lists the relative figure of merit $k_c\sqrt{Q_t}$. Table II lists also $Q_t$ as defined by (4) and $A$ defined by (8). The results of Table II verify that the relative figure of merit gives the same order of coupling efficiency between the cases as $A$. $k_c$, instead, is mostly affected by the size of the transponder coil.

TABLE II
PARAMETER VALUES AND THE SIMULATION RESULTS OF THE TRANSPONDER AND READER COILS AT 867 MHZ.

| $R/\square$ ($\Omega$) | w (mm) | $k_c$ | $Q_t$ | $k_c\sqrt{Q_t}$ | $A$ (dB) |
|---|---|---|---|---|---|
| 5 | 4 | 0.10 | 1.50 | 0.12 | -26.2 |
| 10 | 4 | 0.12 | 0.70 | 0.10 | -28.5 |
| 5 | 7 | 0.13 | 1.84 | 0.18 | -23.1 |
| 10 | 7 | 0.15 | 0.93 | 0.15 | -25.4 |

## III. Fabrication

Fabrication of the transponder includes two steps: printing of the graphene antenna and integration of the antenna and a commercial RF chip Monza R6.

### A. Antenna fabrication

Antenna structures were deposited on flexible substrates by screen printing. Screen printing is a very high volume printing method that can produce thick structured layers with a resolution linewidth of about 100 µm that is sufficient for integration of even small discrete SMD components. Screen printing inks have a high viscosity and solid content, resulting in a thick film after drying. There are many graphene ink providers and the typical achieved conductivity for screen printed graphene-based conductors is 1-10 Ω/□ with 1-3 screen printing passes. The graphene ink used was prepared by TU/e (Eindhoven University of Technology) and is printed as received. This type of printing paste is based on graphene platelets produced by intercalation and thermal expansion in an optimized process for producing a few-layer thick graphene particles suitable for formulation with a variable binder into a viscous dispersion by gelation [13] [14] [15]. Only the graphene platelets contribute to the conductivity and thus the operation of the antenna, and the electrical conductivity of the printed structures used in this work is used as an input for the design modelling. The ink is compatible with plastic substrates and thus the antenna structures were printed on polyethylene terephthalate (PET), polyethylene naphthalate (PEN), polyimide (PI) and paper (Lumisilk) substrates. A laboratory scale printing device (EKRA E2) shown in Fig. 4 was used for printing. The device is equipped with a camera, which allows multilayer printing. The screen used has a mesh of 78 stainless steel threads per cm (200 threads per inch), open area of 47% and a theoretical ink transfer volume of 54 ml/m$^2$. Three layers of graphene were printed with a drying step (100 °C, 10 minutes) in a hot air oven after each printing repetition. For high-throughput printing, a single print layer is desired. Furthermore, multilayer printing increases the risk of shorts in particular where small components are attached, due to mismatch between the layers. The lateral rigidity of the substrate also determines how small structures can be printed in multilayers, since thermal relaxation occurs in most of the flexible thin film materials. The printing layout contains a range of dimensional variation for the antenna structures, with a varying gap for the chip integration and varying antenna structure dimensions, the most sensitive being the conctact pads where the chip is attached; the contact dimensions and the gap between the contacts is less than 200 µm and the chip size less than 500 µm. In order to further increase the conductivity of the printed graphene, the PI samples were annealed at up to 320 °C for 36 minutes. Due to the good conductivity results of thermal annealing, the PI samples were used for the final chip integration. For the rest of the substrates that are more temperature-sensitive, photonic annealing could be used [5]. The annealing removes some of the electrically insulating binder from the ink, leaving a higher fraction of conductive graphene in the printed structure. However, after annealing, the layer is porous due to the binder removal and the sample was compressed using a compression roller system (4 passes, 80 bar, 2 m/min) [15]. The conductivity increases significantly with each process step; the triple-layer printed structure has a sheet resistance of 135 Ω/□ while the annealed and compressed structure has a sheet resistance of < 10 Ω/□, which is a suitable conductivity level for the antenna modelling, thus justifying the extra processing steps. The fabricated prototypes of graphene near field UHF RFID antennas with different dimensioning are shown in Fig. 5.

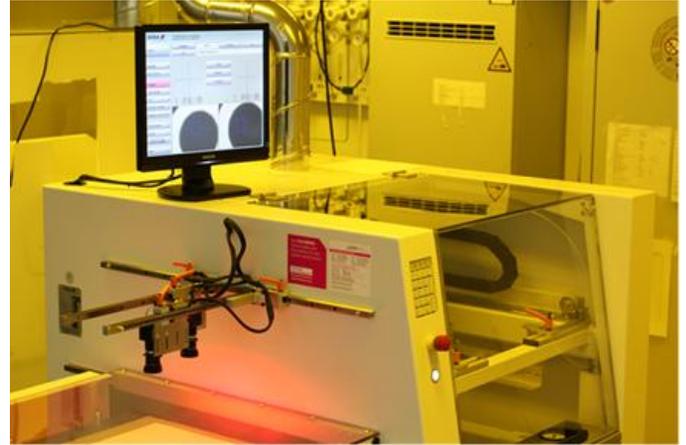

Fig. 4. EKRA S2 screen printing device.

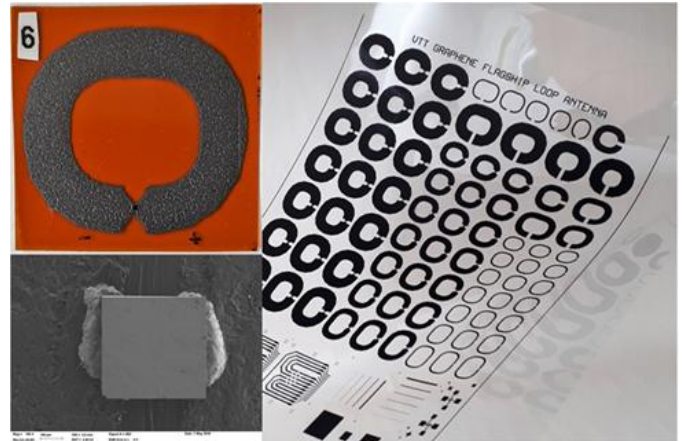

Fig. 5. Printed antenna structures (right) and the attached bare die chip (left). The SEM picture shows the integration of the chip using isotropic conductive adhesive.

### B. Antenna and RF chip integration

A challenge of combining any graphene-based elements, in our case antennas, with conventional electronics is about forming an electrical connection between two dissimilar materials: printed graphene and metal. In former publications about implementing RFID transponders with graphene-based antennas, this has been typically solved by using an aluminum- or copper-based strap that is glued with conductive silver-based adhesive to the graphene antenna. [4] [5]. The problem with such an implementation is that the process needs more steps and individual parts than the conventional way of producing an RFID transponder. Consequently, it does not bring the whole potential of cost-savings and eco-friendliness of a graphene-based transponder into action. Therefore, in this work, the goal was to develop a process based on conductive adhesives for attaching the RFID microchip into a graphene-based antenna. For the bonding process, rough and non-uniform surfaces of printed graphene is a challenge. High and anisotropic resistivity of graphene also increases the contact resistance as the contact area of the microchip is small.

The assembly of chips for RFID transponders in high volumes is usually based on the use of adhesives: anisotropic conductive adhesive (ACA) or Isotropic conductive adhesives (ICA). The ACA contains small amounts of conductive sphere particles dispersed in a polymer matrix. Therefore, the ACA is not conductive in the horizontal plane, eliminating the risk of short-circuits. The processing is started by dispensing the adhesive onto the chip area. Then the chip is aligned with the pads of the substrate and pressed down at a certain temperature. The temperature depends on the substrate and the adhesive. The curing time is also short, typically ~10 s. The ACA suits well to mass production and for fine-pitch applications. The thermocompression of chips is typically done in a batch process, which further speeds up the assembly.

ICAs, made up of a composition of polymer resin and conductive silver fillers, provide electrical conductivity in X, Y and Z directions when the silver flakes pack together due to the shrinkage of the ICA after being cured [16]. ICA allows higher conductivity due to higher filling load of highly conductive Ag flake-shaped particles, but they request wider pitching for chips. ICAs are preferable for applications in which pitches are not critical but higher conductivity of the contact is required.

The electrical resistance of the adhesives is in a range of 4 $\mu\Omega * m$. For comparison, the electrical resistance of typical solder materials, such as PbSn or SnAgCu with different compositions, is in the range of ~0.1 $\mu\Omega * m$. Although the electrical resistance is higher in the case of adhesives, they provide other benefits over solder materials. For example, the processing temperature is lower than that of solders. Additionally, the maximum curing temperature can be tuned, which enables the use of variety of substrate materials. For example, the melting temperatures of eutectic PbSn and SnAgCu are 183 and 217-220 °C, respectively. In practice, the maximum temperature during soldering is typically 30-50 °C higher than the melting temperature. The final properties of adhesives are achieved by curing (i.e. by cross-linking of polymer chains). Different adhesives are cured in different ways, although heat is the most common one. For example, the curing temperature for H20E can vary from 80 to 175 °C. The duration is then varied from 3 h to 45 s, respectively. Due to these benefits, conductive adhesive was selected for this case. Today, chip assembly with ACA on PET substrate is the standard technology for mass production of RFID tags and the use of ACA on e.g. paper substrate has also been successfully demonstrated [16] [17].

Monza R6 chip has a rectangular shape with the outer dimensions of 464.1 µm x 442 µm. The chip has two metal pads for antenna connection. The size of the pads is 166 µm x 422 µm with a spacing of 112 µm between them. The bonding pads are large enough, allowing the use of ICA instead of ACA for better electrical contact between a chip and a graphene antenna.

The isotropically conductive adhesive Epotek H20-E was manually dispensed onto the pads of the chip. The chip was then flip-chipped onto the graphene antenna with Finetech Fineplacer flip-chip bonder. The antenna printed on polyimide (PI) was kept at the room temperature. The chip was aligned to the proper area of the antenna using a camera alignment system. The alignment accuracy with the bonder is in the range of a few micrometers, which was adequate in this case. The assembled transponder was then placed into a drying oven at the temperature of 100 °C for 30 min. The adhesive changes into electrically conductive during this drying step. The obtained connection between the chip and the graphene antenna is mechanically strong. A photograph of the assembled module is shown in Fig. 6.

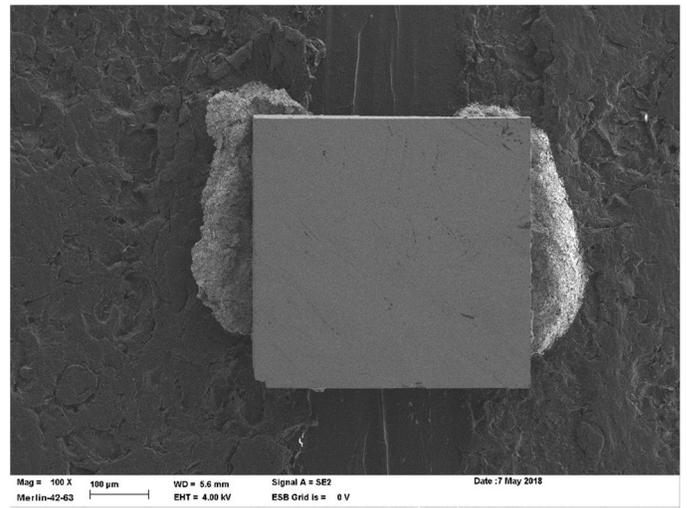

Fig. 6. Monza R6 chip bonded by ICA glue to the printed graphene antenna.

In order to evaluate the feasibility and the electrical characteristics of the different ways of attaching the microchip to the antenna, prototypes were made with three attachment methods: an aluminum strap glued with silver lacquer, the microchip attached directly with silver lacquer and the microchip attached with ICA glue.

## IV. RESULTS AND DISCUSSION

The fabricated tag prototypes were measured using Voyantic Tagformance™ UHF RFID characterization system with an Impinj Matchbox near field reader antenna [10] [11]. The measurement setup inside the anechoic cabinet of the system is shown in Fig. 7. Even though echoes are not that critical for a near-field system, the Faraday-cage shielding from external signals is important. On the Impinj Matchbox antenna (black) there is a 10 mm spacer of Styrofoam (blue) to maintain a constant distance between the measurement antenna and the tag under test. The measurement setup is similar to the simulation model shown in Fig. 1. In the measurement, the activation level of the transponder is measured as a function of frequency [18] [19]. Although in an inductive near field system the maximum reader power is not restricted similarly to radiating systems, the maximum power specification of the reader antenna sets the limit to +30 dBm (1 W) of conducted power [10]. Therefore, in the following, the sensitivity of the prototypes is evaluated in terms of the power marginal that is the difference between the needed activation level and the maximum allowed +30 dBm. In order to compare the simulation and measurement results, the power needs to be defined similarly for the two. This is achieved by defining the 50-Ohm input port of the antenna as the reference point of the RF power. In the simulations, the power marginal $A_M$ is then defined in dB form:

$$A_M(dB) = F + P_{tx\,max} - P_{sens\,IC}, \quad (9)$$

where $F$ is the power coupling coefficient as defined by (6) and (7), $P_{tx\,max}$ is the maximum RF power (+30 dBm) and $P_{sens\,IC}$ is the wakeup (read) sensitivity of the microchip, -20 dBm for Monza R6 [12]. The corresponding power marginal can be calculated from the measurement results:

$$A_M(dB) = P_{tx\,max} + L_{circ} + L_{cable} - P_{tx}, \quad (10)$$

where $L_{circ}$ is the attenuation of the circulator between the RF output of the Tagformance device and the Matchbox antenna and $L_{cable}$ is the attenuation of the coaxial cables and connectors between the circulator and the Matchbox antenna. $P_{tx}$ is the RF output power of the Tagformance device that is recorded as a function of frequency during the measurement. The value of $L_{circ}$, as given by the specification of the device, is -0.2 dB and that of $L_{cable}$, obtained by a measurement, is -1.4 dB.

The different prototypes fabricated and measured are shown in Fig. 8. Two conductor widths, 7 mm and 4 mm, and three methods of attaching the microchip to the coil: strap with silver lacquer, chip directly attached with the same lacquer and chip directly attached with ICA, have been studied. The prototype on the top left has a 7 mm wide conductor and a strap-attached microchip (series "7 mm strap"). The prototype on the top right has a 7 mm wide conductor and a microchip attached directly with silver lacquer ("7 mm silver"). The one on the bottom left has a 4 mm wide conductor and a microchip attached with a strap ("4 mm strap"). The one on the bottom right has a 4 mm wide conductor and a chip attached with ICA ("4 mm ICA"). The outer dimensions of the tags in mm are 27 * 24.5 ("7 mm strap"), 29.5 * 24.5 ("7 mm silver"), 21 * 18.5 ("4 mm strap") and 21 * 18.1 ("4 mm ICA").

Fig. 9 shows the measured power marginal $A_M$ as a function of frequency for the transponders of Fig. 8 at the distance of 10 mm from the surface of Matchbox reader coil. The operation band specified for Matchbox is 865…956 MHz, which should be taken into account when analyzing the results. The measurement result of a commercial aluminum-based near field UHF RFID transponder, "UPM Trap" is also included for comparison [20]. The dashed lines represent the simulation results for the 4 mm wide conductor with two values of square resistance of graphene: 5 Ohms and 10 Ohms. The challenge of the electromagnetic modelling of graphene is about printed graphene being an anisotropic material, the conductivity of which is higher in the horizontal direction. The DC measurement made on the surface of the printed graphene has showed square resistance values slightly lower than 10 Ohms, but as the material defined in the simulation model is homogeneous and isotropic, the two values have been studied by simulations. Another thing that is difficult to model is the contact resistance that depends on the attachment method of the microchip. A strap with silver lacquer forms a big contact area (2 * 3.5 mm$^2$) with a lower series resistance, whereas the direct gluing with ICA forces the RF current to flow through very small area (166 μm x 422 μm) of high-resistivity graphene, resulting in high contact resistance. In terms of the contact resistance, the direct attachment with silver lacquer (case "7 mm silver") lies between these two. The anisotropic nature of graphene relates also to the contact resistance, since, depending on the attachment method, the portion of the current flowing in the vertical direction varies. Geometrically, the simulation model is closest to the ICA bonded case.

Based on the maximum variation in a repeatability test with ten measurements for each case, with the transponder repositioned between the measurements, the repeatability is +/- 0.5 dB over the whole frequency range. According to measurements, moving the tag horizontally 5 mm from the middle produces a maximum attenuation of 1.0 dB (+/- 0.5 dB) within the 865…956 MHz operation band of Matchbox.

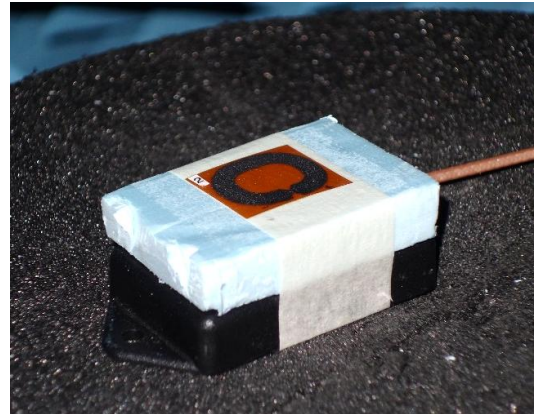

Fig. 7. Measurement setup with an Impinj Matchbox near field reader antenna, Styrofoam spacer and a tag under test inside an anechoic cabinet.

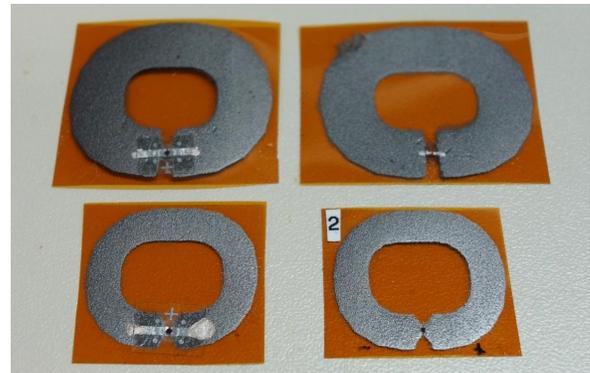

Fig. 8. Near field UHF RFID tag prototypes on graphene: with a 7 mm wide conductor and the microchip attached by a strap of aluminum (top left), with a 7 mm wide conductor and the microchip attached with silver lacquer (top right), with a 4 mm wide conductor and the microchip attached with a strap of aluminum (bottom left) and with a 4 mm wide conductor and the microchip attached with isotropic conductive adhesive (ICA) (bottom right).

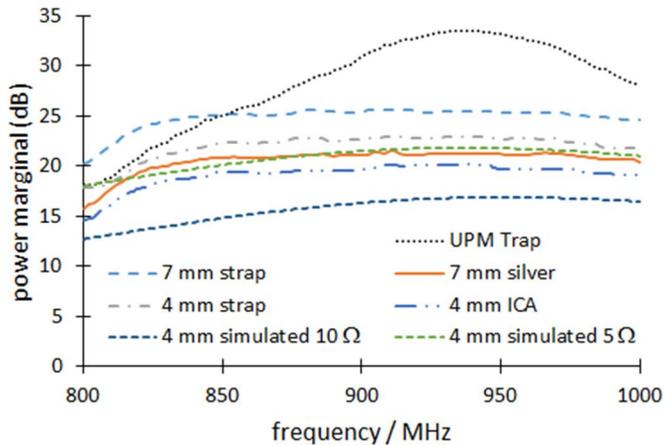

Fig. 9. Power marginal of the transponders of Fig. 8 and a commercial near field tag "UPM Trap" as a function of frequency at the distance of 10 mm from the surface of Matchbox reader coil.

One can conclude from Fig. 9 that there is a sufficient power marginal available for all of the tags measured. Comparing the graphs "7 mm strap" and "4 mm strap" shows that the wide conductor gives better sensitivity due to lower resistive losses, as expected. The comparison of "7 mm strap" with "7 mm silver" shows the difference between the attachment methods, as does the comparison between "4 mm strap" and "4 mm ICA". The results confirm that, in terms of the increasing contact resistance, the order of the attachment methods is strap, silver lacquer and ICA. However, there is still a good power marginal of 19 dB at the frequency of 867 MHz with the ICA-bonded prototype. The curves of the magnetic field intensity as a function of distance for Matchbox predict this marginal to equal the read range of about 40 mm from the surface of Matchbox, whereas for a typical patch antenna the equivalent read range is about 90 mm [10].

Of the prototypes with a 4 mm wide conductor, the one with the ICA contact is between the lines of the simulation results, showing that if the effective square resistance would be adjusted to match the measurement result, the value would be between 5 Ohms and 10 Ohms, but closer to 5 Ohms. The measurement curves bending differently to the simulated ones close to the 800 MHz end may be related to the limited bandwidth of Matchbox. The commercial transponder, "UPM Trap" shows larger marginal that is better sensitivity than the graphene prototypes, even though it uses a less sensitive microchip (Monza 3). The shape of its curve also shows higher Q value than any of the graphene transponders, which was expected due to its aluminum conductor.

Additionally to a simple and straightforward structure, the size of an antenna is an important parameter when aiming at low-cost production. A UHF near field transponder combines these two properties. The size benefit of a near field UHF transponder is especially true for graphene-based transponders, for which implementing an efficient far field antenna requires a very large conductor area. [4] [5] This is due to the low conductivity of graphene, which also affects the performance and size of a near field transponder. However, the conductor area of a near field transponder as well as the absolute size remain moderate. Even though the read range of a near field transponder is clearly shorter than that of a far field tag, a short and well-defined read area is a benefit in many applications. The combination of a short read range and a small, low-cost transponder makes the graphene-based near field transponder an ideal solution for item-level tagging of small objects.

## V. Conclusion

Near field UHF RFID transponders with a screen-printed graphene antenna were designed, fabricated and tested. All of the transponder prototypes fabricated were proven operational with a fair power marginal. The fabrication method, particularly the bonding of a chip to a printed graphene antenna is based on already-existing industrial processes, which makes the concept viable also commercially. The approach demonstrated here encompasses an industrially relevant printing technique, post-processing methods compatible with roll-to-roll manufacturing paradigms and temperature sensitive flexible substrates.

## Acknowledgment

Prof. Heiner Friedrich and Dr Artur Pinto are acknowledged for providing the graphene ink and for discussions regarding the printing and the post processing parameters. The authors would like to thank Impinj for providing the Monza R6 microchips. The research leading to these results has received funding from the EC. The authors would like to thank the Directorate-General for Science, Research and Development of the European Commission for support of the research. Authors would like to acknowledge the Graphene Flagship for the support of the research.

## References

[1] K. Finkenzeller, "RFID Handbook: Radio-Frequency Identification Fundamentals and Applications," 2nd ed. New York, NY, USA, Wiley, 2004.
[2] R. Want, "RFID explained: A primer on radio frequency identification technologies", Synthesis Lectures on Mobile and Pervasive Computing, vol. 1, Jan. 2006, pp. 1-94
[3] J. Landt, "The History of RFID", IEEE Potentials, vol. 24, no. 4, Oct.-Nov. 2005, pp. 8 - 11,
[4] M. Akbari, M. Waqas, A. Khan, M. Hasani, T. Björninen, L. Sydänheimo, L. Ukkonen, "Fabrication and characterization of graphene antenna for low-cost and environmentally friendly RFID tags," *IEEE Antennas Wireless Propag. Lett.* vol. 15, pp. 1569–1572, (2016).
[5] K. Arapov, K. Jaakkola, V. Ermolov, G. Bex, E. Rubingh, S. Haque, H. Sandberg, R. Abbel, G. de With, H. Friedrich, "Graphene Screen-printed Radio-frequency Identification Devices on Flexible Substrates", *Physica status solidi (RRL) - Rapid Research Letters,* vol. 10, issue 11, pp. 812-818 (2016)
[6] P. Nikitin, K. V. S. Rao and S. Lazar "An Overview of Near Field UHF RFID," 2007 IEEE International Conference on RFID, Grapevine, TX, USA, March 26-28, (2007).
[7] R. M. Duarte and G. K. Felic, "Analysis of the Coupling Coefficient in Inductive Energy Transfer Systems," *Active and Passive Electronic Components*, Vol. 2014, Article ID 951624.
[8] M. Kesler, "Highly Resonant Wireless Power Transfer: Safe, Efficient, and over Distance," WiTricity Corporation, 2017, accessed on Jan 6, 2019. [Online]. Available: http://witricity.com/wp-content/uploads/2016/12/White_Paper_20161218.pdf

[9] N. Nitta, N. Li, K. Fujimori and Y. Sugimoto, "A Q-enhancement method for resonant inductive coupling wireless power transfer system," 2017 IEEE Wireless Power Transfer Conference (WPTC), 2017.
[10] Impinj Matchbox™ Near Field UHF RFID Antenna, accessed on Oct 14, 2018. [Online]. Available: https://support.impinj.com/hc/en-us/articles/202755658-MatchBox-Antenna-Datasheet
[11] Y. S. Ong, X. Qing, C. K. Goh, Z. N. Chen, "A segmented loop antenna for UHF near-field RFID", 2010 IEEE Antennas and Propagation Society International Symposium, 11-17 July 2010, doi: 10.1109/APS.2010.5561018
[12] Datasheet of Impinj Monza R6 UHF RFID Microchip, accessed on Oct 14, 2018. [Online]. Available: https://support.impinj.com/hc/article_attachments/115001963950/Monza%20R6%20Tag%20Chip%20Datasheet%20R5%2020170901.pdf
[13] K. Arapov, A. Goryachev, G. de With, and H. Friedrich, "A simple and flexible route to large-area conductive transparent graphene thin-films," *Synthetic Metals*, vol. 201, pp. 67–75, 2015.
[14] K. Arapov, E. Rubingh, R. Abbel, J. Laven, G. de With, and H. Friedrich, "Conductive screen printing inks by gelation of graphene dispersions," *Advanced Functional Materials*, vol. 26, no. 4, pp. 586–593, 2016.
[15] K. Arapov, G. Bex, R. Hendriks, E. Rubingh, R. Abbel, G. de With, and H. Friedrich, "Conductivity Enhancement of Binder-Based Graphene Inks by Photonic Annealing and Subsequent Compression Rolling," *Advanced Engineering Materials*, vol. 18, no. 7, pp. 1234–1239, 2016.
[16] H. Chu, B. An, F. Wu and Y. Wu, "RFID Tag Packaging with Anisotropically Conductive Adhesive," Electronic Packaging Technology, 2006. ICEPT '06. 7th International Conference on , pp. 1-4, 26-29 Aug. 2006.
[17] J. S. Rasul, "Chip on paper technology utilizing anisotropically conductive adhesive for smart label applications," *Microelectronics Reliability,* Vol. 44, No. 1, Jan. 2004, pp. 135-140.
[18] Tagformance by Voyantic, accessed on Oct 14, 2018. [Online]. Available: http://voyantic.com/tagformance
[19] P. Nikitin, K. V. S. Rao, S. Lam. UHF RFID Tag Characterization: Overview and State-of-the-Art, accessed on Oct 14, 2018. [Online]. Available: https://pdfs.semanticscholar.org/4c92/ad48e34cf7ef6a11e7652819cb6cb293d9e2.pdf
[20] UPM Trap near field UHF RFID tag, datasheet available via Yumpu.com, accessed on Oct 14, 2018. [Online]. Available: https://www.yumpu.com/en/document/view/21832041/upm-trap-rfid

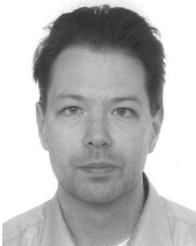
**Kaarle Jaakkola** received the Master of Science (Tech.) degree in electrical engineering from the Helsinki University of Technology (currently Aalto University), Espoo, Finland, in 2003.

Since 2000 he has been working at the VTT Technical Research Centre of Finland, currently as a Senior Scientist. His research interests and expertise include RFID systems, electronics, wireless and applied sensors, antennas, electromagnetic modelling and RF electronics. He has e.g. developed RF parts for RFID systems and designed antennas for both scientific use and commercial products. Antennas designed by him can be found in several commercial RFID transponders. He has authored and coauthored 20 peer-reviewed journal and conference articles and he holds 10 patents as an inventor or a co-inventor.

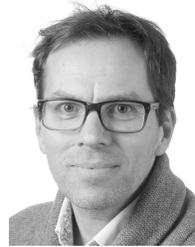
Dr. **Henrik Sandberg** is a principal scientist at VTT's centre for Printable and Hybrid Functionalities. His work is currently focused on printed electronic devices and circuits, polymer device physics, printing technology and printing ink formulation, as well as heterogeneous and monolithic integration for flexible electronics. He coordinates work on printed graphene based materials and hybrid integration, specifically targeting flexible applications such as wearable devices. He specializes in device and circuit development as well as the development of analog printing compatible processing techniques from the lab to the R2R pilot scale and on related ink development, in particular on the topics of thin film polymer transistors and circuits, organic photovoltaics and graphene and other 2D material applications.

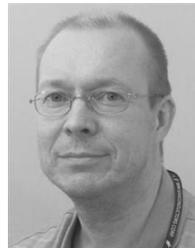
**Markku Lahti** received his M.Sc. and D.Sc. degrees in Electrical Engineering from University of Oulu in 1993 and 2008, respectively. He has been with VTT Technical Research Centre of Finland, Ltd. since 2001, where he is currently working as a Senior Scientist. His research interests are related to the manufacturing of low-temperature co-fired ceramic (LTCC) substrates, and module-level integration and packaging of components with ceramic and polymer platforms. He has been working in several EU and ESA projects, and is author or co-author of over 20 peer-reviewed publications and over 70 conference papers.

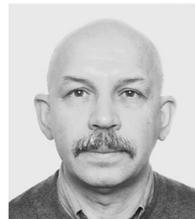
**Vladimir Ermolov** graduated with honours in 1981 and received the Ph.D in 1986 from the Moscow Engineering Physics University (MEPhI). Since 1981, he had been as Senior Research Associate with the laboratory of Dielectric Devices (MEPhI). He worked many times as a visiting researcher in the Department of Physics, Helsinki University and Fraunhofer Institute of Non-destructive testing, Germany.

He had been a principal research scientist with Nokia research center from 1998 till 2011, where he worked as a project manager for many projects in areas of MEMS, nanotechnology, radiotechnology, mass memory technology and sensors. He was involved in commercialization of several technologies. Since 2011 he has been with VTT Technical Research Center of Finland as a principal research scientist, where he had made project management and research in areas of radiotechnology, MEMS, nanotechnology, and technology commercialization.

He is the author and co-author of 54 in referred journals and the holder and co-holder of 47 patents and patents pending. Vladimir Ermolov is a recipient of the 47th European Microwave Conference Microwave Prize.